\documentstyle[12pt]{article}
\begin{document}
\baselineskip=24ptplus.5ptminus.2pt
\vspace*{0.5 in}
\large

\begin{center}
{\bf Phenomenological Local Potentials for $\pi^-$  +  $^{12}$C Scattering
from 120 to 766 MeV } 
\end{center}
\normalsize
\vspace{0.5cm}

\begin{center}
S. W. Hong$^{\ast \dagger}$ and B. T. Kim$^\ast$   \\  
\parskip 2ex
$^\ast$Department of Physics and Institute of Basic Science,
Sungkyunkwan University \\
Suwon 440-746, Republic of Korea  \\
$^\dagger$TRIUMF, 4004 Wesbrook Mall, Vancouver, B.C. Canada, V6T 2A3 
\end{center}

\vspace{0.5cm}
\begin{center}
{\bf Abstract}
\end{center}

\baselineskip=24ptplus.5ptminus.2pt

Pion-nucleus scattering cross sections are calculated by solving
a Schr\"{o}dinger equation reduced from the Klein-Gordon equation. 
Local potentials are assumed, and phenomenological potential
parameters are searched energy-dependently
for $\pi^{-} + ^{12}$C system 
so as to reproduce not only elastic differential cross sections but also 
total elastic, reaction and total cross sections at 13 pion incident
energies from 120 to 766 MeV.
The real and imaginary parts of the local potentials
thus obtained are shown to satisfy the dispersion relation.
The imaginary part of the potentials as a function of the pion
energy  is found to peak near the $\Delta$(1232)-resonance energy. 
The strong absorption radius of the pion projectile
with incident energies near the $\Delta$-resonance region is
found to be about $1.6 A^{1/3}$ fm, which is consistent with previous
studies of the region where the decay of the $\Delta$'s 
takes place in nuclei.
The phenomenological local potentials are then compared with the
local potentials exactly phase-shift equivalent to Kisslinger
potentials for pion energies near the $\Delta$-resonance.

\vspace{1.5em}
PACS number: 24.10.Ht, 25.40.Ny, 25.80.Dj 
\pagebreak

\section{Introduction}

For the last two decades, pion-nucleus elastic scattering has been studied 
by a number of people using a variety of theoretical methods, 
some of which are given in Refs. [$1 - 16$]. The aim of most theoretical 
works is to understand the pion-nucleus scattering and thereby the 
nuclear structure in the framework of the multiple scattering theory
by using elementary pion-nucleon scattering amplitudes obtained
from the pion-nucleon scattering data. Normally, the Klein-Gordon 
equation is solved either in coordinate space \cite{Miller,SMC} or 
in momentum space \cite{Tabakin,Landau} with various forms of elementary 
t-matrices including nuclear structure effects [$1 - 12$]. There are also
works using approaches other than directly solving the Klein-Gordon
equation, such as the $\Delta$-isobar model \cite{Hirata,Karaoglu,Osterfeld} 
taking into account production and propagation of the $\Delta$-isobar 
in nuclei. Though numerous works, essentially solving the 
Klein-Gordon equation in different ways, have revealed much of 
pion-nucleus dynamics, the understanding of the pion-nucleus scattering 
is still not quite satisfactory. 

On the other hand, when one tries to do a quantitative calculation of
cross sections for a reaction involving pions, such as ($\pi$,$\pi'$) 
or ($\pi$,$K$), by using the distorted wave Born approximation 
(DWBA) or the distorted wave impulse approximation (DWIA), 
it is essential to have the pion distorted wave functions. 
Even if the accurate pion distorted wave functions may not be
achieved, it would be convenient to have a way of treating
the distortion effects in a simple manner.
The present study was motivated by such a need of the distorted wave
functions that could reproduce the pion scattering data by 
all means so that the distortion effects could be treated.

The optical potentials normally
used in the Klein-Gordon equation for the 
pion-nucleus scattering are known to be nonlocal, particularly
in the $\Delta$-resonance region due to the $P$-wave nature of the resonance. 
However, it would be not only easier to visualize but also 
interesting if one can localize the nonlocal potential and look at the
dynamics of the scattering from a different point of view.
Recently, Satchler \cite{Satchler} showed a method of reducing 
the Klein-Gordon equation into the form of a Schr\"{o}dinger
equation by redefining some kinematical quantities.
He then  reproduced  not only elastic but also inelastic differential
cross sections of the pion near the 
$\Delta$(1232)-resonance energy for various target nuclei ranging from $^{40}$Ca to
$^{208}$Pb by using local potentials of the Woods-Saxon form.
In Section 2 we follow Satchler, reduce the Klein-Gordon equation
to a Schr\"{o}dinger equation, and search for
phenomenological local optical potentials
which can reproduce the scattering data.

We choose the system of $\pi^- +  ^{12}$C,
because for this system many experimental data are available 
in a wide range of energies, 120 - 766 MeV \cite{Binon,Takahashi,Kahrimanis}
and we are interested in examining the energy-dependency of the  
pion-nucleus local potentials.
We employ the Woods-Saxon form of the local potential \cite{Satchler}
and search for the potential parameters to fit the experimental data.
The real and imaginary parts of the resulting local potentials are
found to be consistent with the dispersion relation.
The energy-dependency
of the phenomenological local potential shows that the imaginary part 
of the potential peaks near the $\Delta$-resonance energy, and it can be 
explained by the decay of the $\Delta$'s in the nuclear medium, 
which is manifested 
as absorption of the pion projectile in the scattering.

We then compare in Section 3 our phenomenological
local potential with the local potential exactly phase-shift
equivalent to the Kisslinger potential
obtained by using
the Krell-Ericson transformation method \cite{Krell},
which has been used for instance by Parija \cite{Parija} and
recently by Johnson and Satchler \cite{JS}.
Section 4 summarizes the paper.

\section{Phenomenological Local Potentials }

\subsection{The Model}

In most works dealing  with solving the Klein-Gordon equation, a so-called 
truncated Klein-Gordon equation is used, which
means terms quadratic in the potentials are neglected compared to 
the pion energy.   Then one gets
\begin{equation}
\left [ -(\hbar c)^2 \nabla ^2 + 2 \omega (V_N + V_C ) \right ] \phi = 
(\hbar kc)^2 \phi ,
\label{eq:eq01}
\end{equation}
where $\phi$ is the distorted wave function for the relative motion between 
the pion and the target nucleus, $V_C$ and $V_N$ are the Coulomb
and the nuclear potentials, respectively, and   
$k$ is the relativistic center-of-mass momentum of the pion.
In regarding Eq. (\ref{eq:eq01}) as the equation for the scattering
between the pion and the target nucleus, 
Stricker, McManus and Carr \cite{SMC} defined $\omega$ by
\begin{equation}
\omega  = \frac{ M_{\pi} m_T c^2 } { M_{\pi} + m_T } ,
\label{eq:SMCEp}  
\end{equation}
where $m_T$ is the target mass given by the target mass number multiplied 
by the
atomic mass unit and $M_{\pi}$ is the total energy of the pion in the 
pion-nucleus center-of-mass system. 
Satchler then introduced a reduced mass $\mu$ defined by $\mu = \omega / c^2 $
and put Eq. (\ref{eq:eq01}) into the form of a Schr\"{o}dinger equation
\begin{equation}
\left [ -\frac{\hbar^2} {2\mu} \nabla ^2 + V_N + V_C \right ] \phi = E_{c.m.} ~\phi
\label{eq:Schr2}
\end{equation}
for the scattering of 
two masses, $M_\pi$ and $m_T$ with a center-of-mass kinetic
energy $ E_{c.m.} = (\hbar k)^2 / 2\mu $. (The incident pion
bombarding energy was modified so that a standard
nonrelativistic optical model computer program could be used \cite{Satchler}.)
In what follows, we use this method of solving Eq. (\ref{eq:Schr2})
with phenomenological Woods-Saxon local potentials.

\subsection{Results for Phenomenological Local Potentials}

The Woods-Saxon form of $V_N$ in Eq. (\ref{eq:Schr2})
can be written as
\begin{equation}
V_N (r) =  \frac{V}{1+exp(X_V ) } + i \frac{W}{1+exp(X_W ) }
\label{eq:ws1}
\end{equation}
with 
\begin{center}
$ X_i = (r-R_i )/a_i , \;\;\;\;\;\;
R_i = r_i A^{1/3}   \;\;\;\;  (i=V, W) , $ \\
\end{center}
where $r_i $ and $a_i $ are radius and diffuseness parameters, respectively,
and $A$ is the target mass number. The Coulomb potential $V_C$ is 
given in a simple form obtained from a uniform
charge distribution of radius $R_C = 1.2 A^{1/3}$ fm.
There are 6 adjustable parameters in Eq. (\ref{eq:ws1}). 
We fixed them by using a $\chi ^2$-fitting 
method. The $\chi ^2$, written explicitly as
\begin{equation}
\chi ^2 = \frac{1}{N} \sum_{i=1}^{N} \left [
\frac{ \sigma_{ex}^i - \sigma_{th}^i }
{\Delta \sigma_{ex}^i } \right ]^2   
\label{eq:eq11}
\end{equation}
is evaluated at each energy, and the potential parameters are adjusted
so as to minimize the $\chi ^2$.
In Eq. (\ref{eq:eq11}), $\sigma_{ex}^i$'s ($\sigma_{th}^i$'s) and
$\Delta \sigma_{ex}^i$'s are the experimental (theoretical)
cross sections and uncertainties, respectively, and $N$ is the 
number of data used in the fitting. Since experimental total elastic
($\sigma_E$), reaction ($\sigma_R$) and total ($\sigma_T$)
cross sections were also available at all energies except for 400 and 500 MeV,
we used not only differential cross sections but also
$\sigma_E$, $\sigma_R$ and $\sigma_T$ as the data,
$\sigma_{ex}^i$, to be fitted. 

In searching for the parameters we first kept
the radius parameters, $r_i$, as 0.9 or 1.0 fm and
the diffuseness parameters, $a_i$, as 0.4 or 0.5 fm. When we could
not get good fits to the data with these fixed parameters, 
we let them vary.
We tried to find both repulsive and attractive potentials
at all energies.
It was possible to find attractive potentials at
all energies considered here, but repulsive potentials were
obtained only at 230, 260 and 280 MeV, which are just above the
$\Delta$-resonance energy.
Satchler also obtained a repulsive Woods-Saxon potential
for $\pi^{\pm}$ + $^{208}$Pb at 291 MeV,
whereas at other (lower) energies he found
attractive potentials \cite{Satchler}.
At these 3 energies the cross sections calculated with attractive
potentials were indistinguishable from
the ones calculated with repulsive potentials.
However, one may normally expect a repulsive potential above the resonance.
Also, the dispersion relation calculations to be discussed in the next
subsection prefer repulsive real potentials just above
the $\Delta$-resonance energy.
Thus, we shall henceforth
include in our discussion only repulsive potentials
at 230, 260 and 280 MeV.
(At energies higher than the $\Delta$-resonance
there exist several $N^*$ resonances, but these resonances are not so pronounced
as the $\Delta$, and they overlap with each other due to broad widths.
Indeed, at other energies we could not find repulsive potentials.)

There is a well-known ambiguity in determining the optical potential
parameters \cite{Satchler,Igo,DNR}. Due to the strong absorption 
taking place in the nuclear surface region, different potentials can often
fit the scattering data equally well as long as they have similar values  
near the surface region. 
Actually, we found that when we used only the
elastic differential cross sections as the data to be fitted, 
the extracted potential parameters
could not always reproduce the experimental total elastic ($\sigma_E$),  
reaction ($\sigma_R$) and total ($\sigma_T$) cross sections. 
However, when we included in the fitting $\sigma_E$, $\sigma_R$ and
$\sigma_T$ as the data to be reproduced 
in addition to the differential cross section, the resulting
potential parameters reproduced all the cross sections quite well as
shown in Figs. 1 and 2. 
(In Figs. 1 and 2, the experimental data for the pion kinetic energies,
$E$ = 120 - 280 MeV and $E$ = 486, 584, 663 and 766 MeV 
are taken from Refs. \cite{Binon} and \cite{Takahashi}, respectively.
Recently, differential elastic cross section data at $E$ = 400 and 
500 MeV became available from Ref. \cite{Kahrimanis}.)
The searched parameters and the $\chi ^2$-values
for each energy are listed in Table 1.
The fits to the experimental cross sections are in general very good.
But there is some discrepancy in the differential cross sections
at low bombarding energies, particularly at $E$ = 150 MeV.
At this energy the second minimum of the differential cross section
was not reproduced correctly, and the third maximum was underestimated.
When we tried to reproduce the data at larger angles
by further adjusting the potential parameters, it was possible to fit the
larger angle data, but then the first minimum was not
reproduced at the right angle.
This may be due to that we have used the Woods-Saxon potentials whereas
more realistic potentials such as local potentials phase-shift equivalent to
Kisslinger potentials \cite{Krell,Parija,JS} look considerably different
from the Woods-Saxon form, particularly inside the nucleus,
as will be shown in Section 3. However, we also note that similar
discrepancies between
the calculated and the experimental cross sections at $E$ = 150 MeV
can be seen from Refs. \cite{Osterfeld} and \cite{Auerbach}, in which
the $\Delta$-isobar model and the Kisslinger potential are used, respectively.

\subsection{Dispersion Relation and Discussion}

Because the extracted potentials have an ambiguity as mentioned above, 
it is worthwhile to check whether they satisfy the dispersion 
relation, which is known to be satisfied by the real and imaginary parts of
the optical potentials \cite{DNR}. Also, since the relativistic 
Klein-Gordon equation, normally solved with nonlocal
potentials, is reduced to a nonrelativistic 
Schr\"{o}dinger equation, it would be interesting to
see whether the phenomenological local potentials  
are consistent with the dispersion relation.
The relation is often written in the form of a so-called subtracted dispersion
relation as 
\begin{equation}
V(E,r) = V(E',r) + \frac{ E - E'}{\pi} P 
\int _0 ^ \infty dE''  
\frac{ W(E'',r) } { (E'' - E' ) ( E'' - E ) }, 
\label{eq:dr}
\end{equation}
where $P$ stands for the principal value and $E'$ is the energy where
$V(E=E',r)$ is known \cite{Nagarajan,Fulton,Hong-Kim}.
This equation tells us that    
once the imaginary part of the potential at a certain radius is known
as a function of the energy the
real part can be calculated from the relation. 
Thus, we inserted the imaginary potentials extracted from the $\chi ^2$-fitting
into the $W(E,r)$ of Eq.(\ref{eq:dr}), computed the real potential using 
the relation, and compared the results with the real potentials
extracted from the $\chi ^2$-fitting. In so doing, since the potential 
values in the nuclear surface region are most significant in determining the 
cross section, we first evaluate a strong absorption 
radius($R_S$)\cite{Satchler,DNR} defined as the apsidal distance 
on the Rutherford trajectory corresponding to the angular momentum 
$L = L_{1/2}$, where $L_{1/2}$ is the angular momentum for which the $S$-matrix 
element has the magnitude $|S_L | = \sqrt{1/2}$. Here, we used non-integer 
$L_{1/2}$ values at which $|S_L | = \sqrt{1/2}$, following 
Ref. \cite{Satchler}. The $L_{1/2}$ values thus obtained are listed in 
Table 1. The strong absorption radius parameter ($r_S = R_S / A ^{1/3}$) 
computed from these $L_{1/2}$ values are also listed in Table 1 and
plotted in Fig. 3(a) as a function of the pion energy.
The $r_S$ becomes as large as 1.6 fm near the $\Delta$-resonance region
and about 1.1 fm at $E \approx 400 - 500$ MeV.
Note that Satchler's strong absorption radius parameters were also
roughly around 1.5 fm near the $\Delta$-resonance energy \cite{Satchler}.
To compare the values of the real and imaginary parts of the potentials
at a certain radius, we took $r = 1.5 A^{1/3}$ fm in Eq. (\ref{eq:dr}), which is 
close to a strong absorption radius near the $\Delta$-resonance energy.

In Fig. 4 is plotted by the solid circles the
extracted real and imaginary parts of the potentials 
evaluated at $r = 1.5 A^{1/3}$ fm as a function of the energy. 
We could roughly fit the solid circles in Fig. 4(b)  by the sum of
a Gaussian function and a constant of the form
\begin{equation}
W(E, r) = W_0 \; \; exp( - ( \frac{E-E_0 }{\Delta E} )^2 ) + W_1,
\label{eq:W0}
\end{equation}
where $W_0$, $E_0$, $\Delta E$, and $W_1$ were found 
to be -13.9, 220, 110, and -3.34 MeV, respectively.
The $W(E,r)$ in Eq. (\ref{eq:W0}) with these parameters is plotted
by the curve in Fig. 4(b).
We then inserted Eq. (\ref{eq:W0}) into Eq. (\ref{eq:dr}),
chose the value of $E'$ in Eq. (\ref{eq:dr}) as 500 MeV, carried out   
the integral over the energy, and  
obtained the real part, $V(E,r)$.
The resulting $V(E,r)$ is
plotted by the curve in Fig. 4(a), 
which roughly fits the extracted real potentials (the solid circles). 
As mentioned earlier, at 230, 260 and 280 MeV both attractive and
repulsive potentials could fit the cross section data equally well.
But the $V(E,r)$ calculated from the dispersion relation (the full curve)
prefers repulsive potentials at these energies.

The dispersion relation is applied at other
radii also from $1.3 A^{1/3}$ to $2.0 A^{1/3}$
fm. In this radial region the extracted imaginary potentials 
can be fitted by Eq. (\ref{eq:W0}) (but with different values of 
$W_0, E_0, \Delta E,$ and $W_1$, of course, for each energy)
with $E_0 \approx $ 205 MeV on the average. 
The real potentials calculated from the dispersion
relation in this radial region reproduce the extracted real potentials
just as well as in Fig. 4(a), so the figures are not repeatedly shown here.
But outside this radial region, the extracted imaginary potentials are not so well
fitted by Eq. (\ref{eq:W0}).
Also, the extracted real potentials are somewhat
more scattered around the real potential curves calculated
from the dispersion relation.
Thus, it seems that the extracted phenomenological local potentials
are consistent with the dispersion relation
in this outer surface region. 

Here, we note that although we reproduce 
the cross sections quite well as shown in Figs. 1 and 2 and
the extracted local
potentials in the outer surface region are reasonably 
consistent with the dispersion relation,
it still does not necessarily mean that the extracted
potentials are the unique ones which can describe the data.
Particularly the inner part of the potentials can be quite different,
as will be seen in Section 3,
because this method cannot well determine the potentials inside the
nucleus due to the absorption in the nuclear surface region in addition to
the fact that we have assumed the Woods-Saxon form of the local potential.

However, an interesting feature obtained from the results is 
the broad peak in the imaginary potential as seen in Fig. 4(b).    
For $1.3 A^{1/3}$ fm $< r < 2.0 A^{1/3}$ fm,
the peaks are located at
$E = E_0 \approx $ 205 MeV on the average, 
which is near the $\Delta$-resonance energy,
and the $\Delta E$ of the peaks is about 110 MeV.
The $\Delta$'s produced in  nuclei can get absorbed
via processes such as  quasi-free decay or spreading
\cite{Hirata,Karaoglu,Udagawa1,Udagawa2}.
Such an absorption is reflected in the flux of the incident pion as  
the imaginary part of the local potential peaked at about $ E = E_0$.
The region where the $\Delta$ decays
in nuclei through the quasi-free channel and the spreading was 
studied \cite{Udagawa2}. It was
shown that the quasi-free decay takes place at $r \geq 1.6 A^{1/3}$ fm and
the spreading at $r \geq 0.9 A^{1/3} $ fm.
Since the quasi-free decay is the dominant decay process
over the spreading by a factor of roughly 2 \cite{Udagawa1,Udagawa2},
the strong absorption radius near the $\Delta$-resonance energy is mainly 
determined by the region where the quasi-free decay takes place,
which is as large as about $1.6 A^{1/3}$ fm. 
Fig. 3(a) and Table 1 show just that the extracted strong absorption radius 
parameters near the $\Delta$-resonance are about 1.6 fm, consistent 
with the results of Ref. \cite{Udagawa2}. Here, we also remark that 1.59 times 
$\pi R_S ^2$ roughly  reproduces the total cross section as plotted by
the crosses in Fig. 3(b), {\it i.e.}, $\sigma _T \approx 1.59 \pi  R_S ^2 $,
where $R_S = r_S A^{1/3}$ with $r_S$ being the strong absorption radius parameter 
plotted by the squares in Fig. 3(a).

\section{Comparison with Local Potentials Equivalent to Kisslinger potential}

In this Section we briefly describe the Krell-Ericson 
transformation method \cite{Krell} closely following Johnson and Satchler
\cite{JS}, where it was extensively applied to both
$\pi^+$ and $\pi^-$ scattered from various target nuclei
at pion energies from 20 to 291 MeV.
A detailed study of the resulting equivalent local potentials
can be found there.

We return to the truncated Klein-Gordon equation in Eq. (\ref{eq:eq01}).
$\omega$ is now taken as the total energy of the pion
in the pion-nucleus center-of-mass system. 
These slightly different definitions of $\omega$ 
do not make any significant difference
in the values of $\omega$ because the mass of the
pion is very small compared to that of target nuclei. 
For the nuclear potential $V_N$ in Eq. (\ref{eq:eq01})
the Kisslinger form \cite{Kisslinger}
of the potential has been frequently used,
which can be written as
\begin{equation}
V_N = \frac{ (\hbar c)^2}{2\omega } \left [ q(r) + \vec{\nabla} 
\cdot \alpha (r) \vec{\nabla} \right ],
\label{eq:Kisspot}
\end{equation}
where the first term $q(r)$ is mainly due to the pion-nucleon $S$-wave 
interaction and the second term comes from the $P$-wave part.

By rewriting the pion distorted wave function $\phi ({\bf r})$
in Eq. (\ref{eq:eq01}) as
\begin{equation}
\phi({\bf r})=P(r)\psi({\bf r})
\label{eq:phi}
\end{equation}
with the Perey factor $P(r)= \left [ 1-\alpha(r) \right ] ^{-1/2}$,
one can get a Schr\"{o}dinger
equation for $\psi({\bf r})$,
\begin{equation}
\left [ -\frac{\hbar^2} {2\mu} \nabla ^2 + U_L + V_C \right ] \psi = E_{c.m.} ~\psi,
\label{eq:Schr}
\end{equation}
where $\mu =\omega /c^2$ and $E_{c.m.} = (\hbar k)^2 / 2\mu$. 
Here, $U_L$ is a local potential dependent only on $r$ as follows:
\begin{equation}
U_L = U_1 + U_2 + U_3 + \Delta U_C
\label{eq:ul}
\end{equation}
with
\begin{eqnarray}
U_1 &=&  \frac{(\hbar c)^2} { 2\omega} \frac{q(r)}{1-\alpha(r)}, \nonumber \\
U_2 &=& -\frac{(\hbar c)^2} { 2\omega} \frac{k^2 \alpha(r)}{1-\alpha(r)}, \nonumber \\
U_3 &=& -\frac{(\hbar c)^2} { 2\omega} \left [ \frac{ \frac{1}{2} \nabla ^2
         \alpha(r) } {1-\alpha(r) }  + \{ \frac{ \frac{1}{2} \nabla
         \alpha(r) } {1-\alpha(r) }    \}^2 \right ],
\label{eq:U123} \\
\Delta U_C &=& \frac{ \alpha(r) V_C } {1-\alpha(r)}. \nonumber
\end{eqnarray}
Thus if $q(r)$ and $\alpha(r)$ are given,  
the local potential $U_L$, exactly phase-shift equivalent to the Kisslinger
potential, can be calculated.
The expressions and parameters for
$q(r)$ and $\alpha(r)$ for the system of
$\pi^-$ + $^{12}$C at 7 energies from 120 to 280 MeV
are available from the work of Sternheim and Auerbach \cite{Auerbach},
where simple forms of $q(r)$ and $\alpha(r)$ are used:
$q(r)=b_0 k^2 \rho(r)$ and $\alpha(r)=b_1 \rho(r)$ with $\rho(r)$ being
the target nuclear density.
(Note that $b_1$ here corresponds to $c_0$ in the notation of Johnson and
Satchler \cite{JS}.)
We have used "Fermi averaged parameters"
for $b_0$ and "Fitted parameters" for $b_1$ as listed in Table I of
Ref. \cite{Auerbach}.
The same parameter sets were used by Di Marzio and Amos 
to calculate approximate analytic pion distorted wave functions \cite{Amos}.
We took the $^{12}$C density from Ref. \cite{Amos}, 
which is consistent with Ref. \cite{dejager}.

The real and imaginary parts of
$U_L$ calculated with these parameters 
are plotted by the full curves in Figs. 5(a) and 5(b), respectively,
for $E =$ 120, 150, 230, and 280 MeV only.
(For brevity, figures for 180 and 260 MeV are omitted. $U_L$ for 200 MeV
is shown in Fig. 6.)
Both real and imaginary parts of $U_L$ display a wiggly behaviour, 
as already observed in Refs. \cite{Parija} and \cite{JS},
but the wiggles disappear as the energy increases.
The reason for the gradual disappearance of the wiggles at
higher energies can be easily seen, when the Kisslinger potential
is expressed in a simple form as in Refs. \cite{Auerbach} and \cite{Amos}.
In Fig. 6, each term of $U_L$ is plotted for two different
pion energies; one below the $\Delta$-resonance,
the other above the resonance.
The real and imaginary parts are plotted by the full and broken curves,
respectively. It is easily seen that $U_2$ and
$U_3$ are the terms that characterize the shape of the summed
local potential $U_L$. $U_2$ determines the overall shape of $U_L$
for both below and above the resonance, and $U_3$ brings in more
fluctuations, particularly below the resonance.
As the energy increases the wiggles in all terms of $U_L$
become less prominent.
Such gradual disappearance of oscillatory behaviours of
the real potential at higher energies
was already observed in a model-independent
Fourier-Bessel analysis of the pion potential by Friedman \cite{Friedman},
who extracted the real potential by  assuming the Woods-Saxon
form for the imaginary potential.
The same tendency of disappearance of wiggles at higher energies
can be also seen from the figures in Refs. \cite{Parija} and \cite{JS}.

We can also see that the wiggles do not appear in the outer nuclear
surface region, where 
the scattering is most sensitive to the potential.
Thus, as the energy increases,  
the equivalent local potentials at large radii become more or less 
close to the form of a Woods-Saxon potential.
In Fig. 5 we plotted by the broken curves the
phenomenological Woods-Saxon potentials extracted in Section 2.
At 230, 260 and 280 MeV the phenomenological local potentials are
close to the equivalent local potentials
$U_L$ (the full curves) in the outer surface region.
Especially, the imaginary parts 
at large radii are very close to each other.
But at lower energies there are large discrepancies
between the phenomenological Woods-Saxon potentials
and the equivalent local potentials.
Even the signs of the real potentials are opposite except for large radii
at 120 MeV.
(We may, however, remark that even the equivalent local
potentials could have different signs of real potentials
at smaller radii depending on the 
interaction parameters used as shown in Fig. 13 of Ref. \cite{JS}, 
while they produce similar scattering cross sections, because
the scattering is most sensitive to the potential at large radii.)
As pointed out earlier, we can see from Fig. 4(a)
that the real potentials at
$1.3 A^{1/3}$ fm $ < r < 2.0 A^{1/3}$ fm are repulsive only at
the energies just above the $\Delta$-resonance. Thus,
the dispersion relation calculations seem to require attractive potentials
at energies below the $\Delta$-resonance in the outer surface region.

Also, although the equivalent local potentials are theoretically
better founded, the phenomenological Woods-Saxon potentials reproduce
the cross sections much better, as shown in Fig. 1.
(
The equivalent local potentials obtained here result in the same
differential cross sections as in Figs. 1 and 2 of Ref. \cite{Auerbach},
so they are not  repeated here.)

\section{Summary}

We assumed the Woods-Saxon form of phenomenological local potentials
in solving a Schr\"{o}dinger equation reduced from the Klein-Gordon equation
and searched for the potential parameters.
The calculated
cross sections reproduced the experimental cross sections quite well
in a wide range of energy.
The real and imaginary parts of the phenomenological potentials in the outer
nuclear surface region are found to
satisfy the dispersion relation.
The imaginary part of the phenomenological local potentials
as a function of the energy has
a peak near the $\Delta$-resonance energy due to
the decay of the $\Delta$'s in the nuclear medium, which is 
reflected in the pion flux as absorption of the incident pion.
The strong absorption radius ($ \approx 1.6 A^{1/3}$ fm) 
in the $\Delta$-resonance region is
found to be consistent with the previous studies of the region where
the $\Delta$ decays in the nuclear medium.
But we again stress that the phenomenological local potentials obtained here
are not necessarily unique.
This method of calculating the pion 
cross sections may rather be taken as a simple way of taking into account 
the distortion effects in the DWBA or DWIA calculations 
with a relatively good accuracy as in Ref. \cite{Satchler}. 
It is well known that for high energy beams the distortion
effects can be often treated by an eikonal approximation with 
so-called a distortion factor or an attenuation factor. 
Indeed, Table 1 shows that at higher energies the real part of the
phenomenological local potential becomes much smaller than the
imaginary part.

Very recently this approach to the treatment of the distortion effects has been
applied to the $^{12}$C($\pi^+ , K^+$)$^{12}_{\Lambda}$C reaction, and
the distorted wave functions of $\pi^+$ and $K^+$ calculated in this
method have been successfully used in reproducing the hypernuclear
production cross sections in DWIA \cite{Lambda}.
Even if the distorted wave functions calculated in this way
may not be accurate especially inside the nucleus,
this simple method seems quite
useful in dealing with the distortion effects in view of the fact 
that most of the cross sections are well reproduced. 

\vspace{2pc}\hfill\newline

{\bf Acknowledgements}\hfill\newline
 
We are grateful to Professor T. Udagawa 
for his hospitality at the University of Texas at Austin 
and for his careful reading the manuscript and helpful discussions.
SWH owes thanks to Drs. H. W. Fearing and B. K. Jennings for their hospitality 
at TRIUMF and discussions and Professor B. C. Clark for sending some of 
the experimental data in numbers.
This work was supported in part
by the Ministry of Education of Korea (BSRI 98-2422) 
and by Korea Science and Engineering Foundation 
(951-0202-033-2).

\pagebreak

\renewcommand
\arraystretch{1.8}
\begin{tabular}{c|ccc|ccc|ccc}    \hline\hline

~$E$~ & ~~~~$V$~~~~&~~~$r_V$~~~&$~~a_V~~$&~~~$W$~~~ 
&~$r_W$&~~$a_W$~~&~~$\chi^2$~~&~~$L_{1/2}$~~&~$r_S$~ \\ 
(MeV) &  (MeV) & (fm) & (fm) &(MeV) & (fm) & (fm) &  & ($\hbar$) & (fm) \\ \hline
120 &-31.2~&1.55           &.257  &-149. &\underline{0.9}  &.536 &  1.7   & ~~3.5    &1.58  \\ \hline
150 &-54.0~&1.20           &.574  &-103. &\underline{1.0} &.566 &  8.9   & ~~4.1    &1.58  \\ \hline
180 &-86.4~&\underline{0.9}&.553  &~-62.0&1.30 &.455 &  2.7   & ~~4.7    &1.59  \\ \hline
200 &-93.6~&\underline{0.9}&.571  &~-66.35&1.20 &.4905 &  2.0   & ~~5.0    &1.54  \\ \hline
230 & 137.~&\underline{1.0}&.2156 &-58.53 &1.40 &.3556 &  1.8   & ~~5.1  &  1.44  \\ \hline
260 & 111.~&\underline{1.0}&.3136 &-53.8 &1.35 &.3694 &  1.2   & ~~5.3   &  1.37  \\ \hline
280 & 109.~&\underline{1.0}&.319  &-46.74 &1.35 &.381 &  0.63   & ~~5.5  &  1.34  \\ \hline
400 &-39.4~&1.12           &\underline{.4}  &~-59.8&\underline{0.9}&.474 &  1.5   & ~~5.7    &1.05  \\ \hline
486 &-22.0~&1.124          &\underline{.4}  &~-70.3&\underline{0.9}&\underline{.4}&  1.8 &~~7.0 &1.11 \\ \hline
500 &-31.8~&1.05           &\underline{.4}  &~-53.8&\underline{0.9}&.540 &  1.7   & ~~6.7    &1.05  \\ \hline
584 &-12.4~&1.20           &.366  &~-69.8&\underline{0.9} &.437 &  1.3   & ~~8.3    &1.13  \\ \hline
663 &-4.90~&1.37           &.300  &~-64.0&0.965&.442 &  1.9   & ~~9.6   &1.17  \\ \hline
766 &-4.11~&1.40           &.526  &~-60.3&\underline{1.0} &.462 &  1.8   & 11.6    &1.25  \\ \hline\hline
\end{tabular}

\vspace{1cm}

\begin{table}
\baselineskip=24ptplus.5ptminus.2pt

\caption{The $\pi^{-} + ^{12}$C phenomenological local optical potential
parameters obtained
from the $\chi ^2$-fitting. The parameters are fixed such that they 
can reproduce not only differential elastic cross sections 
but also total elastic ($\sigma_E$), reaction ($\sigma_R$)
and total ($\sigma_T$) cross sections at all energies except
for 400 and 500 MeV, where only differential cross sections
are available. The radius and/or diffuseness parameters
fixed during the search are underlined.
The extracted $L_{1/2}$ and
strong absorption radius parameters are also listed at each energy.
In computing the $\chi^2$-values for $E$ = 486, 584, 663 and 766 MeV,
only the systematic errors of the experimental differential cross sections
were taken for $\Delta \sigma_{ex}^i$.}


\end{table}

\pagebreak

\begin{figure}
\caption{
The calculated differential elastic cross sections (the full curves)
are compared with the experimental data for $\pi^- + ^{12}$C.
The experimental data are taken from Ref. [17] for $E =$ 120 $-$ 280 MeV. 
The data for $E =$ 486, 584, 663 and 766 MeV are from Ref. [18].
For $E =$ 400 and 500 MeV we used the data from Ref. [19].
}
\end{figure}
 
\begin{figure}
\caption{The calculated total elastic($\sigma_E$), reaction($\sigma_R$)
and total($\sigma_T$) cross sections denoted by the crosses
are compared with the experimental cross sections for $\pi^- + ^{12}$C.
The data are from Refs. [17] and [18].}
\end{figure}

\begin{figure}
\caption{(a) The strong absorption radius parameters ($r_S$)
as a function of the incident pion energy.
(b) $1.59 \pi  R_S ^2$ denoted
by the crosses are compared with the experimental total cross sections. 
The data are taken from Refs. [17] and [18].}

\end{figure}

\begin{figure}
\caption{The real and imaginary potentials  extracted from 
the $\chi ^2$-fitting are computed at $r = 1.5 A ^{1/3}$ fm 
and are plotted by the solid circles in (a) and (b), respectively.
The extracted imaginary potentials (the solid circles in (b)) are fitted 
by the curve as explained
in the text. The curve in (a) is the real potential  
calculated from the dispersion relation.}
\end{figure}

\begin{figure}
\caption{The local potentials $U_L$ exactly phase-shift equivalent to
Kisslinger potential calculated by using the Krell-Ericson
transformation for $\pi^- + ^{12}$C system for pion 
energies from 120 to 280 MeV are plotted by the full curves.
(The potentials for 180, 200, and 260 MeV are not shown here for brevity.)
The broken curves are the phenomenological Woods-Saxon potentials.
The real and imaginary parts of the potentials are plotted in
the columns (a) and (b), respectively.
}
\end{figure}

\begin{figure}
\caption{The $U_1$, $U_2$, $U_3$, and $\Delta U_C$ components
and the summed local potential $U_L$ at 120 and 200 MeV
are shown in column (a) and (b), respectively. The full and broken curves
denote the real and imaginary parts, respectively.}
\end{figure}

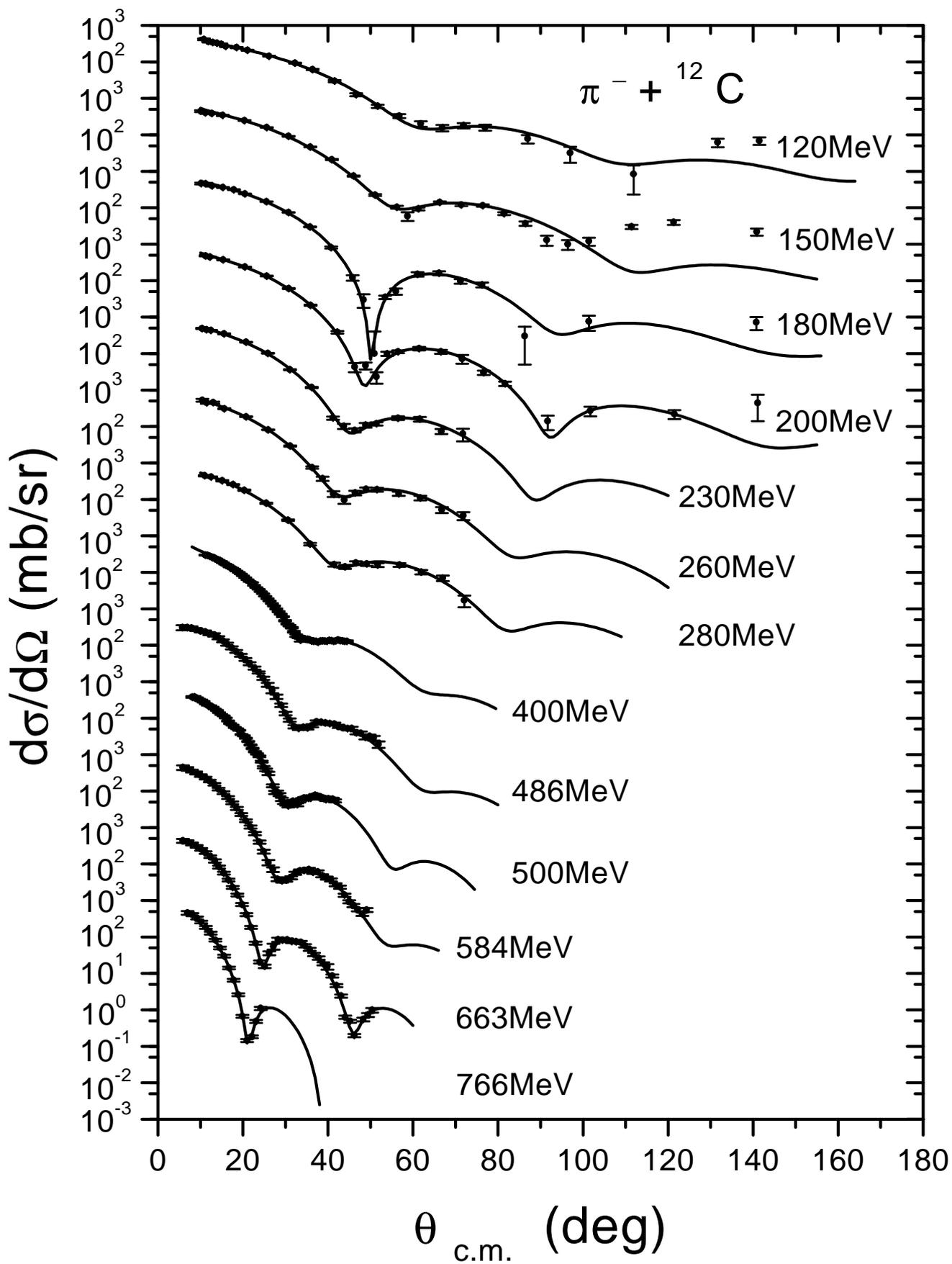
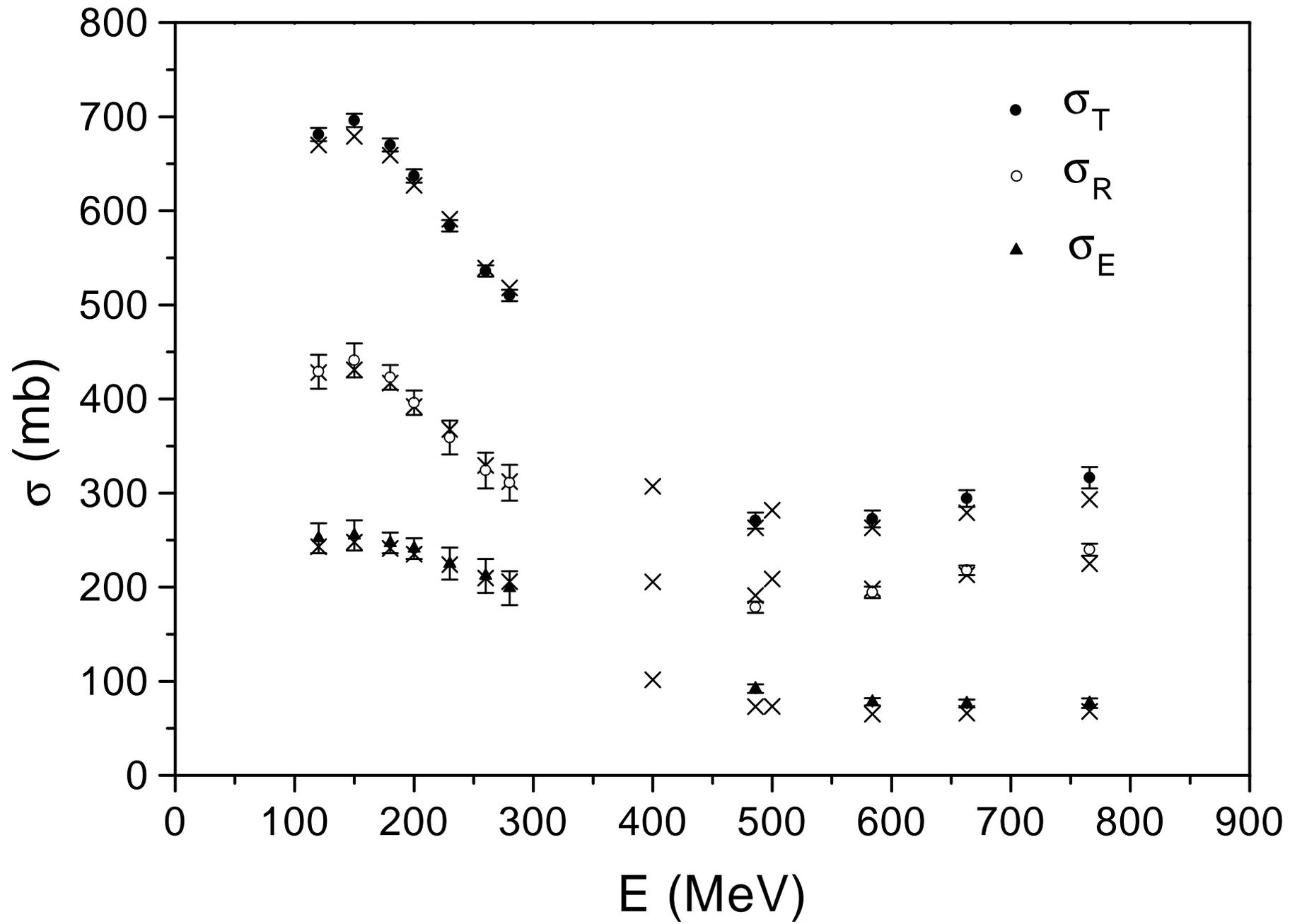
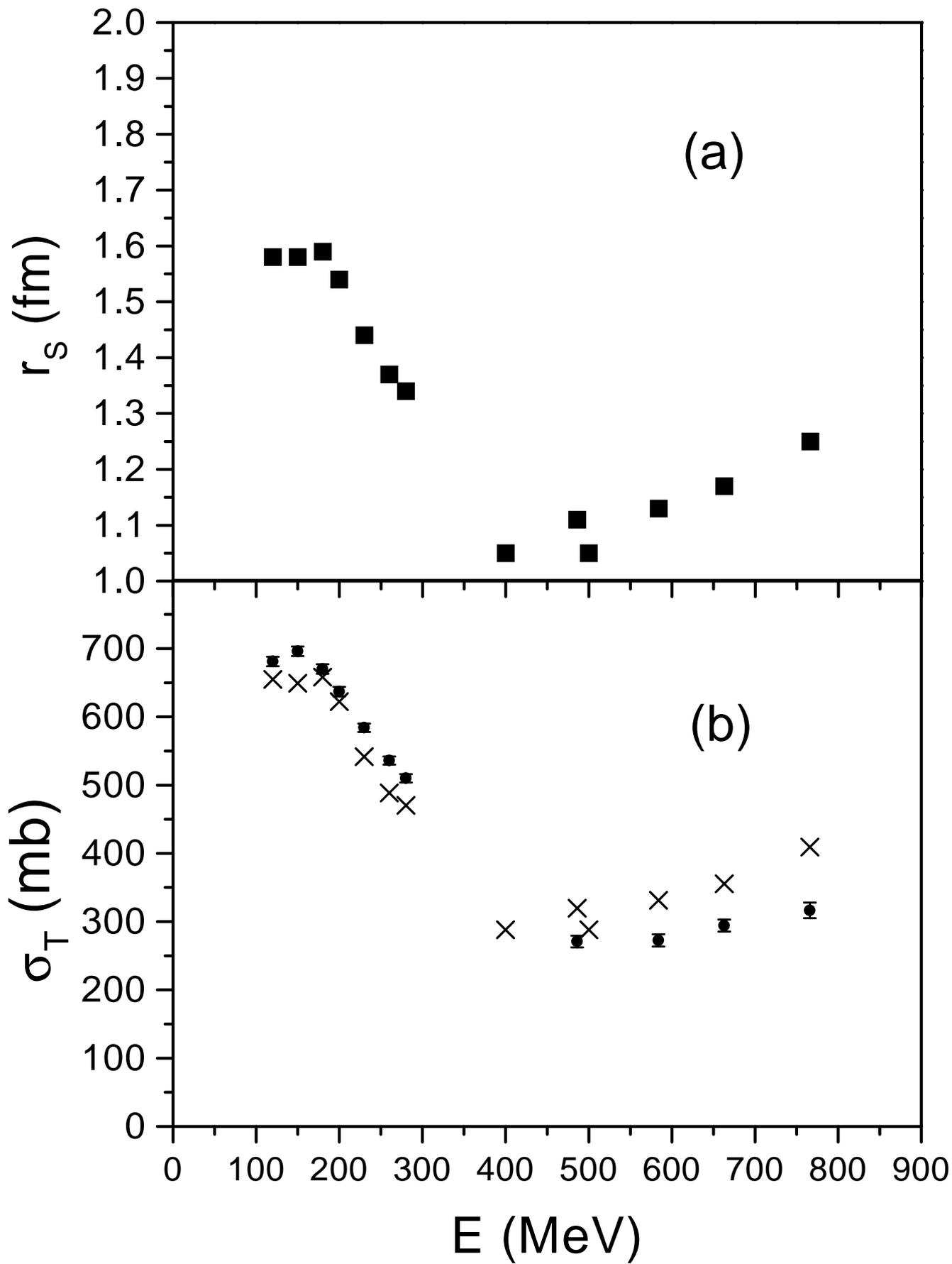
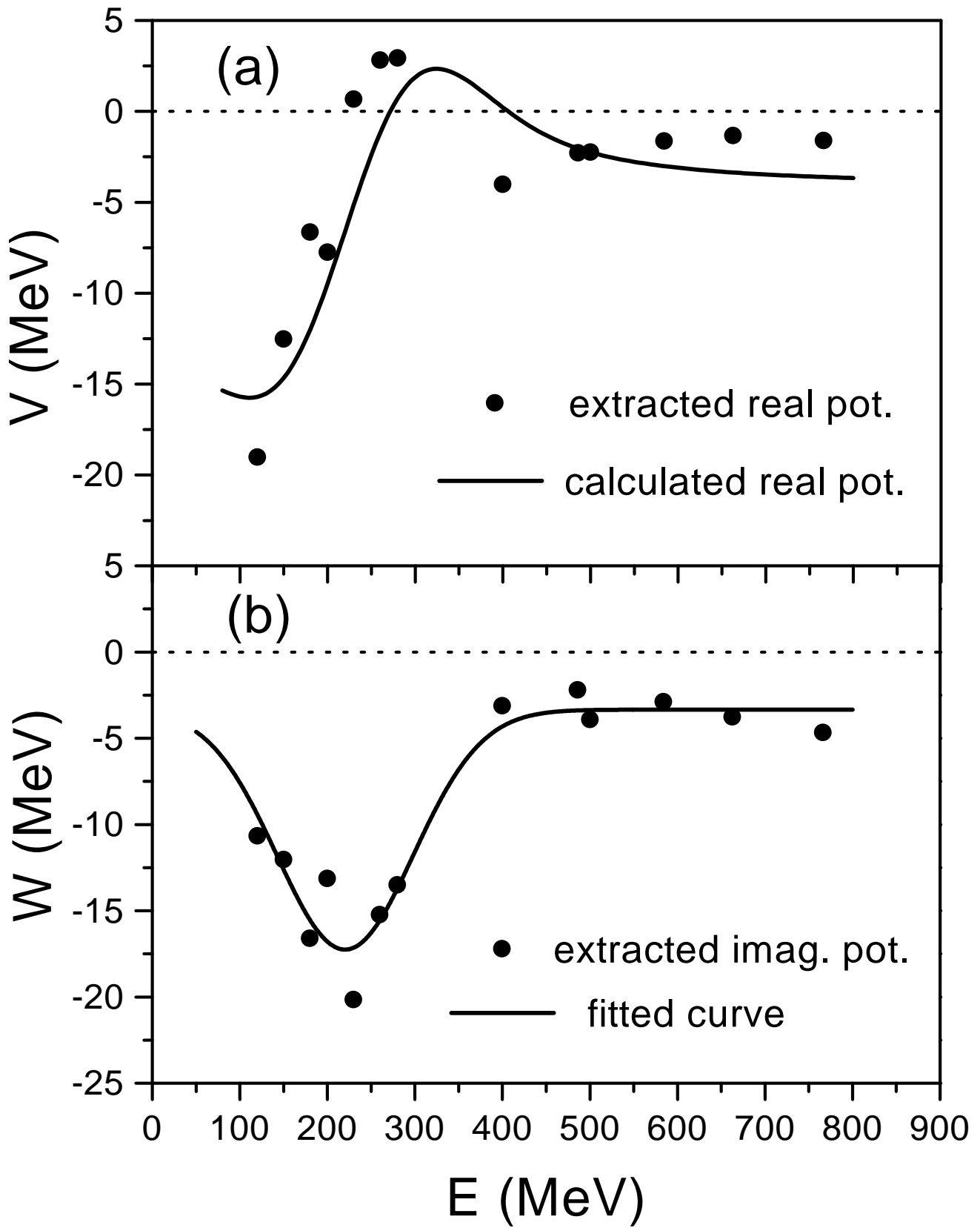
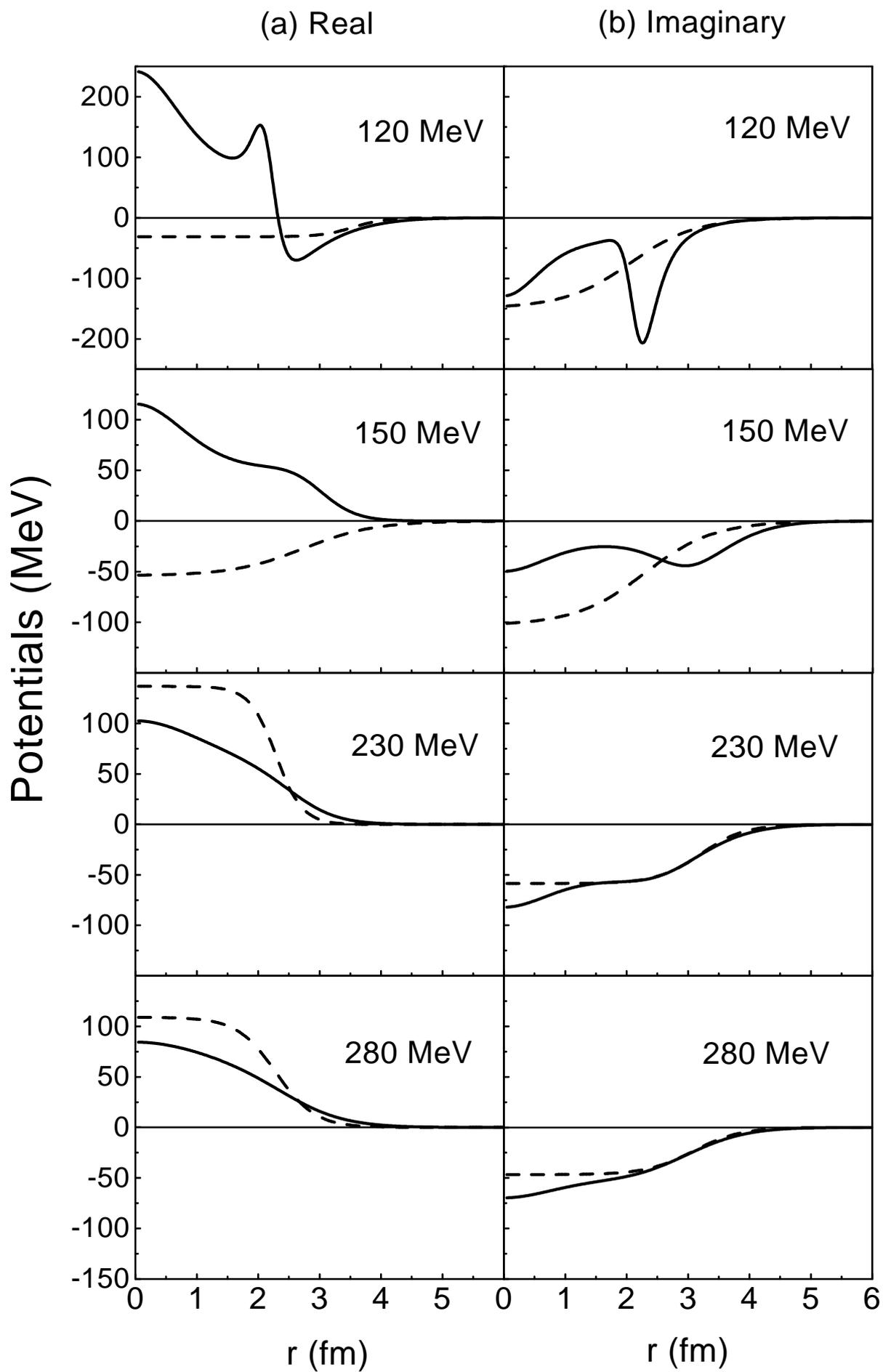
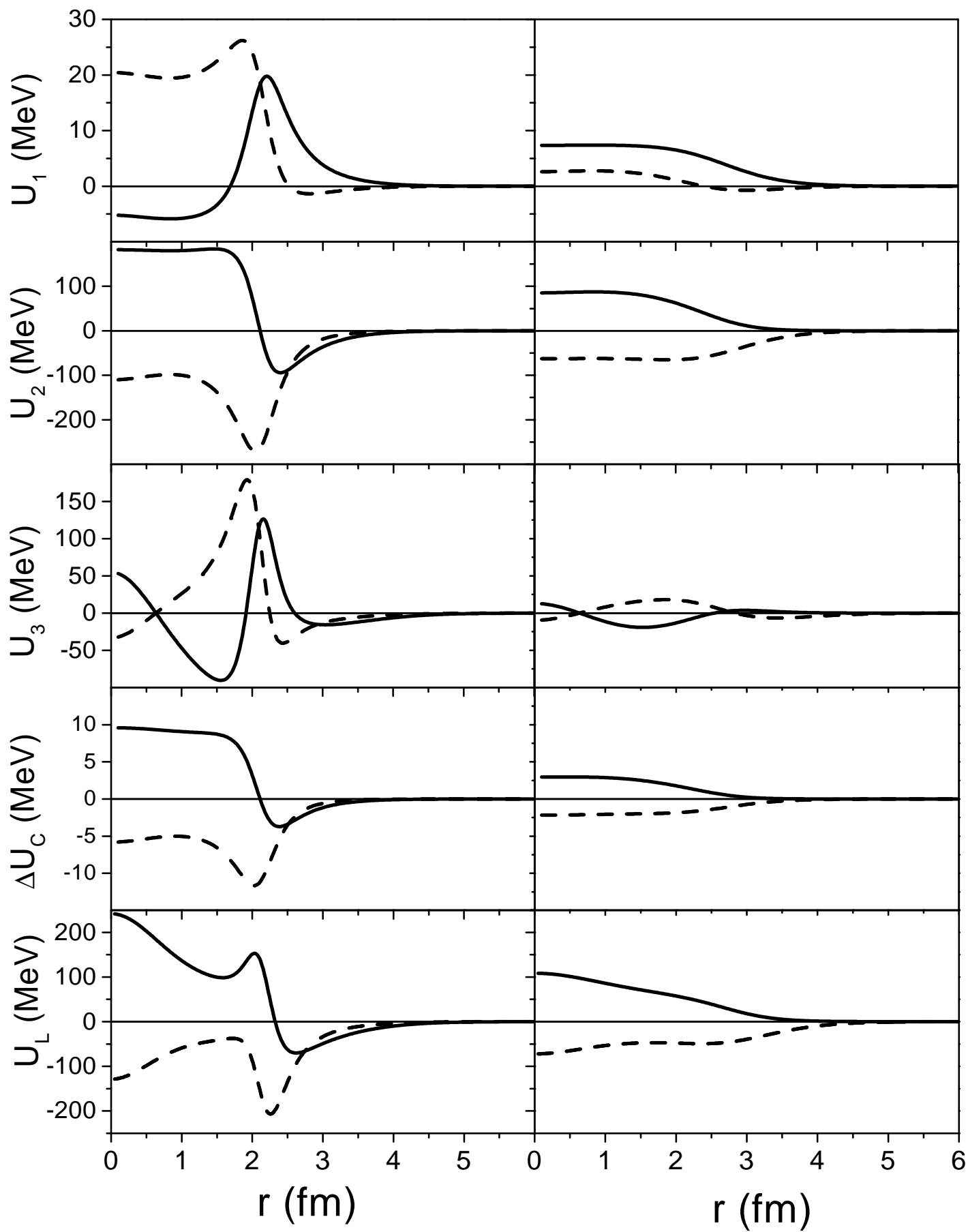
\end{document}